%Paper: hep-th/9209122
%From: T.Eguchi@newton.cam.ac.uk
%Date: Tue, 29 Sep 92 16:45 BST

%
\input phyzzx
%title page
%
\date{September 1992}
\rightline{Newton Institute Preprint NI92004}
\titlepage
\vskip 1cm
\title{$W_\infty$ Algebra in Two-Dimensional Black Hole}
\author {Tohru Eguchi}
\address{Newton Institute for Mathematical Sciences,
Cambridge CB3 0EH, England  \break
and \break Department of Physics, Faculty of Science, University of
Tokyo, Tokyo 113, Japan}
\author{Hiroaki Kanno}
\address{DAMTP, University of Cambridge,
Cambridge CB3 9EW, England}
\andauthor{Sung-Kil Yang}
\address{Institute of Physics, University of Tsukuba, Tsukuba,
Ibaraki 305, Japan }
\abstract{ We study the $SL(2;R)/U(1)$ coset model of two-dimensional
black hole and its relation to the Liouville theory coupled
to the $c=1$ matter. We uncover a basic isomorphism in the algebraic
structures of these theories and show that the black hole model
has the same physical spectrum as the $c=1$ model, i.e.
tachyons, $W_{\infty}$ currents and the ground ring elements.
We also identify the operator responsible for the creation of the
mass of the black hole.}
\endpage

\overfullrule=0pt
%
%references
%
\def\cmp#1{{\it Comm. Math. Phys.} {\bf #1}}

\def\pl#1{{\it Phys. Lett.} {\bf B#1}}

\def\prd#1{{\it Phys. Rev.} {\bf D#1}}

\def\np#1{{\it Nucl. Phys.} {\bf B#1}}

\def\jmath#1{{\it J. Math. Phys.} {\bf #1}}
\def\mpl#1{{\it Mod. Phys. Lett.} {\bf A#1}}

\def\del{\partial}
%%%%%%%%%%%%%%%%%%%%%%%%%%%%%%%%%%%%%%%%%%%%%%%%%%%%%%%%%%%%%%%%%%%%%
In this article we discuss the $SL(2;R)/U(1)$
coset model of two-dimensional black hole
\REF\WITA{E. Witten, \prd{44} (1991) 314.}
[\WITA] and its relation to the Liouville
theory coupled to $c=1$ matter. Making use of free field realization of
$SL(2;R)$ current algebra we first show that the stress tensor of
the coset theory is identical to that of the $c=1$ Liouville theory
up to BRST exact terms and thus there exists a basic isomorphism
in the algebraic structure between these two theories. We next
determine the physical spectrum (BRST invariants) of the black hole theory
which turns out to be identical to that of the $c=1$ two-dimensional gravity.
Besides the standard tachyon states, there also exist discrete states
in two-dimensional black hole which possess the same algebraic structure,
the $W_\infty$-algebra and ground ring,
as in the $c=1$ theory. Thus the black hole and
$c=1$ theory in fact seem to originate from the identical
conformal field theory.
We then try to identify the
operator which is responsible for the creation of the mass of the black hole.
We shall show that the mass perturbation
operator is given by $\beta \exp(-2\sqrt 2\phi)$ $(\beta,\gamma,\phi$
are the free fields used to realize $SL(2;R)$ current algebra),
the screening operator of $SL(2;R)$ algebra.  	In the Liouville
theory the perturbing operator is given by
the cosmological constant $\exp(-\sqrt 2\phi)$
and thus the black hole and $c=1$ gravity theories are the same conformal
thoeries
but perturbed by different marginal operators.
\par
Let us start from the free field realization of $SL(2;R)$ current algebra
\REF\WAKI{M. Wakimoto, \cmp{104} (1986) 605.}
[\WAKI] given by
%%%%%%%%%%%%%%%%%%%%%%%%%%%%%%%%%%%%%%%%%%%%%%%%%%%%%%%%%%%%%%%%%%
$$
\eqalignno{
J_+~=&~\beta~,& (1) \cr
J_-~=&~\beta\gamma^2+\sqrt{2k'}\gamma\del\phi + k\del\gamma~,\quad
k'=k-2~,& (2) \cr
J_3~=&~-\beta\gamma - \sqrt{k'\over2}\del\phi~.& (3) \cr}
$$
%%%%%%%%%%%%%%%%%%%%%%%%%%%%%%%%%%%%%%%%%%%%%%%%%%%%%%%%%%%%%%%%%%
Here $\beta,\gamma$ are the commuting ghost fields with the dimensions $h=1,0$
and we have the operator product expansions (OPE's)
$\beta(z)\gamma(w)\sim 1/( z-w), \phi(z)\phi(w)\sim -log(z-w)$.
$k$ is the level of the $SL(2;R)$ current algebra,
%%%%%%%%%%%%%%%%%%%%%%%%%%%%%%%%%%%%%%%%%%%%%%%%%%%%%%%%%%%%%%%%%%%%%
$$
\eqalignno{
J_+(z)J_-(w) \sim& {k\over (z-k)^2} - {2J_3(w)\over z-w} + \cdots~,
&(4) \cr
J_3(z)J_\pm(w) \sim& \pm {J_\pm(w)\over z-w} + \cdots~, &(5) \cr
J_3(z)J_3(w) \sim& {-k/2\over (z-k)^2} + \cdots~. &(6) \cr}
$$
%%%%%%%%%%%%%%%%%%%%%%%%%%%%%%%%%%%%%%%%%%%%%%%%%%%%%%%%%%%%%%%%%%
Then the Sugawara construction of the stress tensor reads
%%%%%%%%%%%%%%%%%%%%%%%%%%%%%%%%%%%%%%%%%%%%%%%%%%%%%%%%%%%%%%%%%%
$$
\eqalignno{
T_{SL(2;R)}(z) &= {1\over k-2}~\{ {1\over2}J_+(z)J_-(z) +
{1\over2}J_-(z)J_+(z) - J_3(z)J_3(z) \} &(7) \cr
&=~\beta\del\gamma - {1\over 2} (\del\phi)^2 -
{1\over\sqrt{2k'}}\del^2\phi~. &(8) \cr}
$$
%%%%%%%%%%%%%%%%%%%%%%%%%%%%%%%%%%%%%%%%%%%%%%%%%%%%%%%%%%%%%%%%%%%%%
\par
Let us next gauge away the $U(1)$ degrees of freedom
 and construct a coset theory
$SL(2;R)/U(1)$ making use of the BRST method.
We introduce a gauge boson $X$ with $X(z)X(w)\sim -log(z-w)$ and
form a BRST operator for the $U(1)$ symmetry,
%%%%%%%%%%%%%%%%%%%%%%%%%%%%%%%%%%%%%%%%%%%%%%%%%%%%%%%%%%%%%%%%%%%%
$$
Q_{U(1)}~=~\oint dz \xi(z) \big( J_3(z) - i\sqrt{k\over 2} \del X(z)
\big)~.\eqno{(9)}
$$
%%%%%%%%%%%%%%%%%%%%%%%%%%%%%%%%%%%%%%%%%%%%%%%%%%%%%%%%%%%%%%%%%%%%%%
Here $\xi,\eta$ are anti-commuting ghosts with $h=0,1$, respectively, and obey
$\xi(z)\eta(w) \sim 1/( z-w)$. Since the total $J_3$ current,
$J_3^{tot} = J_3 - i \sqrt{k\over 2} \del X$, has no central term,
%%%%%%%%%%%%%%%%%%%%%%%%%%%%%%%%%%%%%%%%%%%%%%%%%%%%%%%%%%%%%%%%%%%%%%%
$$
J_3^{tot}(z)J_3^{tot}(w) \sim \hbox{ regular}~, \eqno{(10)}
$$
%%%%%%%%%%%%%%%%%%%%%%%%%%%%%%%%%%%%%%%%%%%%%%%%%%%%%%%%%%%%%%%%%%%%%
$Q_{U(1)}$ is nilpotent
%%%%%%%%%%%%%%%%%%%%%%%%%%%%%%%%%%%%%%%%%%%%%%%%%%%%%%%%%%%%%
$$
Q_{U(1)}^2 = 0~. \eqno{(11)}
$$
%%%%%%%%%%%%%%%%%%%%%%%%%%%%%%%%%%%%%%%%%%%%%%%%%%%%%%%%%%%%%%%%%%%
The  process of gauging has introduced the new fields $X,\xi,\eta$ and
then the total stress tensor is given by
%%%%%%%%%%%%%%%%%%%%%%%%%%%%%%%%%%%%%%%%%%%%%%%%%%%%%%%%%%%%%%%%%%%%
$$
T_{SL(2;R)/U(1)}~= \beta\del\gamma - \eta\del\xi -{1\over2}(\del\phi)^2
-{1\over\sqrt{2k'}} \del^2\phi - {1\over2}(\del X)^2~.
\eqno{(12)}
$$
%%%%%%%%%%%%%%%%%%%%%%%%%%%%%%%%%%%%%%%%%%%%%%%%%%%%%%%%%%%%%%%%%%%%
When we choose $k =9/4$ or $k'= 1/4$,
$T_{SL(2;R)/U(1)}$ has the central charge $c=26$ and describes a critical
string theory.
\par
Now let us eliminate the $\beta,\gamma,\eta,\xi$ fields from (12)
so that the stress tensor has the form of the Liouville field $(\phi)$
coupled with the matter $(X)$. It is easy to derive the following identity
which is of basic importance in our analysis.
%%%%%%%%%%%%%%%%%%%%%%%%%%%%%%%%%%%%%%%%%%%%%%%%%%%%%%%%%%%%%%%%%%%%%
$$
\eqalignno{
T_{SL(2;R)/U(1)}~=&~ -{1\over 2}(\del\phi')^2  -{1\over\sqrt{2k'}}(\del^2\phi')
 -{1\over 2}(\del X')^2 + \big\{ Q_{U(1)}, -\del(log~\gamma)\cdot \eta \big\}&
(13) \cr
\phi'&= \phi + \sqrt{k'\over 2} log~\gamma~, &(14) \cr
X'&= X + i\sqrt{k\over 2} log~\gamma~, &(15) \cr}
$$
%%%%%%%%%%%%%%%%%%%%%%%%%%%%%%%%%%%%%%%%%%%%%%%%%%%%%%%%%%%%%
In (13) $\beta,\gamma,\eta,\xi$ have in fact disappeared from
the stress tensor except for the shift of the fields $\phi, X$ and
the $Q_{U(1)}$-exact term.
We shoud note that the commuting ghosts $\beta,\gamma$ may be bosonized as
%%%%%%%%%%%%%%%%%%%%%%%%%%%%%%%%%%%%%%%%%%%%%%%%%%%%%%%%%%%
$$
\eqalignno{
\beta =& -i\del v \exp(iv-u)~, &(16) \cr
\gamma =& \exp(u-iv)~, &(17) \cr}
$$
%%%%%%%%%%%%%%%%%%%%%%%%%%%%%%%%%%%%%%%%%%%%%%%%%%%%%%%%%%%%
with $u(z)u(w) \sim -log(z-w)~,
v(z)v(w) \sim -log(z-w)$.
Thus $log~\gamma = u-iv$. Note that it is possible to take fractional
or negative powers of $\gamma$ while the  power of
$\beta$ is restricted to positive integers.
\par
Now we consider the critical case $k=9/4$ and further introduce
the diffeomorphism ghosts $b,c$ (with $h=2,-1$),
%%%%%%%%%%%%%%%%%%%%%%%%%%%%%%%%%%%%%%%%%%%%%%%%%%%%%%%%%%%%
$$
T_{SL(2;R)/U(1)}^{tot}~= -{1\over 2}(\del\phi')^2 -{\sqrt 2}
\del^2\phi' -{1\over 2}(\del X')^2 -2b\cdot \del c - \del b\cdot c
+ \big\{ Q_{U(1)}, -(\del log~\gamma)\cdot \eta \big\}
\eqno{(18)}
$$
%%%%%%%%%%%%%%%%%%%%%%%%%%%%%%%%%%%%%%%%%%%%%%%%%%%%%%%%%%%%%
BRST operator for diffeomorphism invariance is defined as usual by
%%%%%%%%%%%%%%%%%%%%%%%%%%%%%%%%%%%%%%%%%%%%%%%%%%%%%%%%%%%%%%
$$
Q_{diff}~=\oint dz c(z) \big( T_{SL(2;R)/U(1)}(z) + {1\over 2}T_{gh}
\big) \eqno{(19)}
$$
%%%%%%%%%%%%%%%%%%%%%%%%%%%%%%%%%%%%%%%%%%%%%%%%%%%%%%%
where $T_{gh} = -2b\cdot \del c - \del b\cdot c$. Note that
$Q_{diff}^2=0, \big\{Q_{diff}, Q_{U(1)}\big\} =0$.
Let us define our BRST operator as the sum
%%%%%%%%%%%%%%%%%%%%%%%%%%%%%%%%%%%%%%%%%%%%%%%%%%%%%%%%%
$$
Q = Q_{diff} + Q_{U(1)}~. \eqno{(20)}
$$
%%%%%%%%%%%%%%%%%%%%%%%%%%%%%%%%%%%%%%%%%%%%%%%%%%%%%%%%%%%%
Then $T_{SL(2;R)/U(1)}^{tot}$ is further rewritten as
%%%%%%%%%%%%%%%%%%%%%%%%%%%%%%%%%%%%%%%%%%%%%%%%%%%%%%%%
$$
T_{SL(2;R)/U(1)}^{tot}~= -{1\over 2}(\del\phi')^2 -{\sqrt 2}
\del^2\phi' -{1\over 2}(\del X')^2 -2b'\cdot \del c - \del b'\cdot c
+ \big\{ Q , -(\del log~\gamma)\cdot \eta \big\}~,
\eqno{(21)}
$$
%%%%%%%%%%%%%%%%%%%%%%%%%%%%%%%%%%%%%%%%%%%%%%%%%%%%%%%%%%%%%%%%%%%
where
%%%%%%%%%%%%%%%%%%%%%%%%%%%%%%%%%%%%%%%%%%%%%%%%%%%%%%%%%%%%
$$
b' = b + \eta~\del~log~\gamma~. \eqno{(22)}
$$
%%%%%%%%%%%%%%%%%%%%%%%%%%%%%%%%%%%%%%%%%%%%%%%%%%%%%%%%%%%%%%
Thus up to a BRST exact term, the stress tensor has exactly
 the same form as in the $c=1$ Liouville theory
%%%%%%%%%%%%%%%%%%%%%%%%%%%%%%%%%%%%%%%%%%%%%%%%%%%%%%%%%%%%%%
$$
T_{c=1}^{tot}~= -{1\over 2}(\del\phi)^2 -{\sqrt 2}
\del^2\phi -{1\over 2}(\del X)^2 -2b\cdot \del c - \del b\cdot c~,
\eqno{(23)}
$$
%%%%%%%%%%%%%%%%%%%%%%%%%%%%%%%%%%%%%%%%%%%%%%%%%%%%%%%%%%%
at the vanishing value of the cosmological constant $(\mu =0)$. The shift of
the fields eq. (14),(15) have a simple
physical interpretation; $\phi'$ and $X'$ are the charge-neutral versions of
$\phi$ and $X$. In fact it is easy to check that
$\phi'$ and $X'$ are neutral with respect to the $U(1)$ charge
$\oint J_3^{tot}(z)dz$ and thus describe the degree of freedom which
live in the coset space $SL(2;R)/U(1)$. We notice that since $\gamma$
does not generate a singularity when contracted with itself,
$\phi',X',b',c$ have exactly
the same OPE as $\phi,X,b,c$. (21), (23) exhibit the basic identity in the
algebraic structure of the gauged WZW and $c=1$ theories.
\par
Let us next turn to the discussion of the physical spectrum of two-dimensioanl
black hole. Physical states are defined as cohomology classes of the
BRST operator,
%%%%%%%%%%%%%%%%%%%%%%%%%%%%%%%%%%%%%%%%%%%%%
$$
Q \vert phys \rangle =0~, \quad  \vert phys \rangle \simeq
\vert phys \rangle + Q \vert any \rangle~, \eqno{(24)}
$$
%%%%%%%%%%%%%%%%%%%%%%%%%%%%%%%%%%%%%%%%%%%%%
where $Q$ is given by (20).
It turns out that the BRST invariant states in
$SL(2;R)/U(1)$ theory can be obtained from the known BRST invariants
in the $c=1$ theory simply by making use of the transformation
 (14) (15),(22) at $k=9/4$,
%%%%%%%%%%%%%%%%%%%%%%%%%%%%%%%%%%%%%%%%%%%%%%%%%%%%%%%%%%%
$$
\eqalign{
\phi' &= \phi + {1\over 2\sqrt 2} log~\gamma~, \cr
X' &= X + i{3\over 2\sqrt2}log~\gamma~, \cr
b' &= b + \eta\del log~\gamma~.}
\eqno{(25)}
$$
%%%%%%%%%%%%%%%%%%%%%%%%%%%%%%%%%%%%%%%%%%%%%
\par
$\underline{Tachyons}$:
\par
In the $c=1$ theory tachyon wave functions are simply given by the vertex
operators
%%%%%%%%%%%%%%%%%%%%%%%%%%%%%%%%%%%%%%%%%%%%%%%%%%%%%%%%%%%%%%
$$
\exp(ip_X X)\exp(p_L\phi)~, \eqno{(26)}
$$
%%%%%%%%%%%%%%%%%%%%%%%%%%%%%%%%%%%%%%%%%%%%%%%%%%%%%%%%%%%
with $p_X,p_L$ satisfying the on-shell condition,
%%%%%%%%%%%%%%%%%%%%%%%%%%%%%%%%%%%%%%%%%%%%%%%%%%%%%
$$
\pm p_X = p_L + {\sqrt 2}~. \eqno{(27)}
$$
%%%%%%%%%%%%%%%%%%%%%%%%%%%%%%%%%%%%%%%%%%%%%%%%%%%%%%%
In the black hole theory tachyons are given by
%%%%%%%%%%%%%%%%%%%%%%%%%%%%%%%%%%%%%%%%%%%%%%%%%%%%%%
$$
\gamma^{{1\over2\sqrt 2}(-3p_X + p_L)}~\exp(ip_X X)\exp(p_L\phi)~.
\eqno{(28)}
$$
%%%%%%%%%%%%%%%%%%%%%%%%%%%%%%%%%%%%%%%%%%%%%%%%%%%%%%%%%%%
(28) is simply the charge-neutral version of (26).
\par
$\underline{W_\infty currents}$:
\par
In the $c=1$ Liouville theory certain special states appear
at discrete values of the momentum
\REF\POL{A.M. Polyakov, \mpl{6} (1991) 635.}
\REF\GKN{D. Gross, I.R. Klebanov and M. Newman, \np{350}
(1991) 621.}
\REF\PK{I.R. Klebanov and A.M. Polyakov, \mpl{6} (1991) 3273.}
\REF\WITB{E. Witten, \np{373} (1992) 187.}
\REF\LZ{B. Lian and G. Zuckerman, \pl{266} (1991) 21.}
\REF\BMP{P. Bouwknegt, J. McCarthy and K. Pilch, \cmp{145} (1992) 541.}
[\POL,\GKN,\PK,\WITB,\LZ,\BMP];
current operators which form $W_\infty$ (wedge) algebra are given by
%%%%%%%%%%%%%%%%%%%%%%%%%%%%%%%%%%%%%%%%%%%%%%%%%%%%%%%%%%%%%%%%%%%%
$$
\eqalignno{
W_{j,j}^{(+)}~=&~\exp(i{\sqrt 2}jX)\exp({\sqrt 2}(j-1)\phi)~,
\quad j= 0, {1\over2}, 1, \cdots & (29) \cr
W_{j,m}^{(+)}~=&~\bigl( \oint dw \exp(-i{\sqrt 2}X(w)) \bigr)^{(j-m)}
W_{j,j}^{(+)}~, \quad -j\leq m \leq j~.  & (30) \cr}
$$
%%%%%%%%%%%%%%%%%%%%%%%%%%%%%%%%%%%%%%%%%%%%%%%%%%%%%%%%%%%%%%
There also exist the ``negative'' current opeartors
%%%%%%%%%%%%%%%%%%%%%%%%%%%%%%%%%%%%%%%%%%%%%%%%%%%%%%%%%%%%%%%%%
$$
\eqalignno{
W_{j,j}^{(-)}~=&~\exp(i{\sqrt 2}jX)\exp(-{\sqrt 2}(j+1)\phi)~,
\quad j= 0, {1\over2}, 1, \cdots & (31) \cr
W_{j,m}^{(-)}~=&~\bigl( \oint dw \exp(-i{\sqrt 2}X(w)) \bigr)^{(j-m)}
W_{j,j}^{(-)}~, \quad -j\leq m \leq j~.  & (32) \cr}
$$
%%%%%%%%%%%%%%%%%%%%%%%%%%%%%%%%%%%%%%%%%%%%%%%%%%%%%%%%%%%%%%
Corresponding operators in the black hole theory are again
obtained by the simple substitution (25). Explicitly some
examples are given by
%%%%%%%%%%%%%%%%%%%%%%%%%%%%%%%%%%%%%%%%%%%%%%%%%%%%%%%%%
$$
\eqalignno{
W_{0,0}^{(+)}~=&W_{0,0}^{(-)} =
\gamma^{-1/2}\exp(-{\sqrt 2}\phi)~,&(33) \cr
W_{1/2,1/2}^{(+)}~=&\gamma^{-1}\exp({i\over\sqrt 2}X)
\exp(-{1\over\sqrt 2}\phi)~,~
W_{1/2,-1/2}^{(+)}~=\gamma^{1/2}\exp({-i\over\sqrt 2}X)
\exp(-{1\over\sqrt 2}\phi)~,&(34) \cr
W_{1,1}^{(+)}~=&\gamma^{-3/2}\exp(i{\sqrt 2}X)~, \quad
W_{1,0}^{(+)}~={i\over\sqrt 2}\big( \del X + {3i\over 2\sqrt 2}
\del\gamma\cdot\gamma^{-1}\big)~,& \cr
W_{1,-1}^{(+)}~=&\gamma^{3/2}\exp(-i{\sqrt 2}X)~,&(35) \cr
W_{1/2,1/2}^{(-)}~=&\gamma^{-3/2}\exp({i\over\sqrt 2}X)
\exp(-{3\over\sqrt 2}\phi)~, \quad
W_{1/2,-1/2}^{(-)}~=\exp({-i\over\sqrt 2}X)
\exp(-{3\over\sqrt 2}\phi)~,&(36) \cr
W_{1,1}^{(-)}~=&\gamma^{-5/2}\exp(i{\sqrt 2}X)\exp(-2\sqrt2 \phi)~, & \cr
W_{1,0}^{(-)}~=&\beta\exp(-2\sqrt2\phi)~\simeq \gamma^{-1}{i\over\sqrt 2}\big(
\del X +
{3i\over 2\sqrt 2} \del\gamma\cdot\gamma^{-1}\big)
\exp(-2\sqrt2\phi)~,& \cr
W_{1,-1}^{(-)}~=&\gamma^{1/2}\exp(-i{\sqrt 2}X)
\exp(-2\sqrt2\phi)~,&(37) \cr}
$$
%%%%%%%%%%%%%%%%%%%%%%%%%%%%%%%%%%%%%%%%%%%%%%%%%%%%%%%%%%%
It is easy to check that these are in fact invariant under
our BRST operator (20) (to be precise, $cW_{j,m}^{(\pm)}$ are the invariant
operators. $W_{j,m}^{(\pm)}$ is BRST invariant up to a total derivative.)
Since $\phi',X'$ have the same OPE as $\phi,X$, our $W^{(+)}$ currents above
also form the same $W_\infty$ algebra as known in the $c=1$ gravity theory.
\par
$\underline{Ground\ ring}$:
\par
In the $c=1$ Liouville theory the ground ring elements appear
as partners of $W_\infty$
currents at a neighboring ghost number [\WITB]. The black hole theory
also possesses the ground ring elements whose generators as given by
%%%%%%%%%%%%%%%%%%%%%%%%%%%%%%%%%%%%%%%%%%%%%%%%%%%%%%%%%%%%%%%%%%%
$$
\eqalignno{
x =& \big\{ (bc - {i\over\sqrt 2}\del X + {1\over\sqrt 2}\del\phi)
\gamma^{-1/2} + (\del\gamma + \eta c\cdot\del\gamma)\gamma^{-3/2}
\big\} \exp({i\over\sqrt 2}X + {1\over\sqrt 2}\phi)~, &(38) \cr
y =& \big\{ (bc + {i\over\sqrt 2}\del X + {1\over\sqrt 2}\del\phi)
\gamma + (-{1\over 2}\del\gamma + \eta c\cdot\del\gamma)
\big\} \exp(-{i\over\sqrt 2}X + {1\over\sqrt 2}\phi)~, &(39) \cr}
$$
%%%%%%%%%%%%%%%%%%%%%%%%%%%%%%%%%%%%%%%%%%%%%%%%%%%%%%%%%%%
It is also easy to check that $x,y$ are invariant under the BRST operator
(20) and their explicit expressions (38) (39) agree exactly with those
obtained from the known expressions of the $c=1$ theory after
substitution (25). Other elements of the ring are
obtained by multiplying the generators (38) (39) and we obtain a
ring isomorphic to ${\bf C} [x,y]$ as in the $c=1$ theory.
\par
What do these results mean on the relation between the gauged WZW and
the $c=1$ gravity theories? They seem to imply that these theories
in fact originate from the common algebraic structure or
conformal field theory but perhaps
perturbed into different directions. In the Liouville theory the
perturbing operator will be the world-sheet cosmological
constant $\exp(-\sqrt2\phi)$ and the effects of the perturbation
on the $W_\infty$ algebra and the ground ring have been studied in detail
\REF\KACH{S. Kachru, \mpl{7} (1992) 1419.}
\REF\BAR{J.L.F. Barb\'on, preprint CERN-TH 6379/91, January,1992.}
\REF\DOTS{Vl.S. Dotsenko, preprint CERN-TH 6502/92, PAR-LPTHE 92-17,
May, 1992}
[\KACH, \BAR,\DOTS]. In the case of the black hole model, however,
the perturbing operator responsible for the non-zero mass of the black hole
has not been clearly identified
(for previous attempts, see ref.
\REF\BEKU{M. Bershadsky and D. Kutasov, \pl{266} (1991) 345.}
\REF\MASH{E. Martinec and S.L. Shatashvili, \np{368} (1992) 338. }
[\BEKU,\MASH]). In our search for the mass
operator we take the following heuristic approach based on the
analysis of a non-linear $\sigma$-model:
we first write down an action which reproduces the stree-tensor of the coset
theory (13)
as
%%%%%%%%%%%%%%%%%%%%%%%%%%%%%%%%%%%%%%%%%%%%%%%%%%%%%%%%%%
$$
S_0~={1\over 8\pi} \int d^2x \big( \del\phi'\bar{\del}\phi' +
\del X' \bar{\del} X' \big) -{1\over 4\pi\sqrt{2k'}} \int
d^2x \sqrt g R^{(2)} \phi'~, \eqno{(40)}
$$
%%%%%%%%%%%%%%%%%%%%%%%%%%%%%%%%%%%%%%%%%%%%%%%%%%%%%%%%
Here $R^{(2)}$ is the scalar curvature of the Riemann surface
and we have dropped the $Q_{U(1)}$-exact term. We then look for an $(1,1)$
operator which, when added to (40), transforms the flat
target-space metric into that of a black hole.
Such an operator must be of the form
$F_{\phi\phi}(\phi',X')\del\phi'\bar{\del}\phi' +
F_{\phi X}(\phi',X')\big(\del\phi'\bar{\del}\phi' + \bar{\del}\phi'\del X'
\big) + F_{XX}(\phi',X')\del X\bar{\del}X'$:
the new target-space metric becomes
$G_{\phi\phi} = 1 + F_{\phi\phi},  G_{\phi X} =  F_{\phi X},
G_{XX} = 1 + F_{XX}$ which would describe a curved space-time.
Thus the perturbing operator $V$ must be linear in the derivative of
the fields $\phi',X'$ (for both left and right components) and this
requirement uniquely determines the operator $V$.
In fact $V$, being a marginal perturbation, is given by the $(1,1)$
form $W_{j,m}^{(+)}(z) W_{j,m}^{(+)}(\bar{z})$ or
$ W_{j,m}^{(-)}(z) W_{j,m}^{(-)}(\bar{z})$.
$W_{j,m}^{(\pm)}$ currents, however, are differential polynomials
in $\phi',X'$ of degree $j^2 - m^2$ (multiplied by a vertex
operator) and hence only $W_{1,0}^{(+)}, W_{1,0}^{(-)}$ are linear
in the derivative of the fields.
Now $W_{1,0}^{(+)}(z) W_{1,0}^{(+)}(\bar{z}) \simeq \del X'
\bar{\del}X'$ and hence is a trivial change of the radius
of $X'$. Therefore $V = W_{1,0}^{(-)}(z) W_{1,0}^{(-)}(\bar{z}) $
is the unique candidate for the black hole mass operator.
As shown in (37), $W_{1,0}^{(-)}$ is just the screening operator of
the $SL(2;R)$ current algebra
%%%%%%%%%%%%%%%%%%%%%%%%%%%%%%%%%%%%%%%%%%%%%%%%%%%%%%%%%%
$$
W_{1,0}^{(-)}=\beta\exp \big(-\sqrt{2\over k'}\phi \big)
\eqno{(41)}
$$
%%%%%%%%%%%%%%%%%%%%%%%%%%%%%%%%%%%%%%%%%%%%%%%%%%%%%%%%%%%
and expected to appear naturally in the theory. It has in fact been
suggested to be the black hole mass operator by Bershadsky and Kutasov
in ref.[\BEKU]. Up to a $Q_{U(1)}$-exact term $W_{1,0}^{(-)}$ may be
written as
%%%%%%%%%%%%%%%%%%%%%%%%%%%%%%%%%%%%%%%%%%%%%%%%%%%%%%%%%%%%%
$$
W_{1,0}^{(-)}~=~-
\bigl( \sqrt{k'\over 2}\del\phi'
+ i\sqrt{k\over 2} \del X' \bigr)
\exp\big( -\sqrt{2\over k'}\phi' \big)
\eqno{(42)}
$$
%%%%%%%%%%%%%%%%%%%%%%%%%%%%%%%%%%%%%%%%%%%%%%%%%%%%%%%%%%%%%%
which we employ for the discussion of the mass perturbation.
Then the perturbed action is given by
%%%%%%%%%%%%%%%%%%%%%%%%%%%%%%%%%%%%%%%%%%%%%%%%%%%%%%%%%%%%
$$
\eqalign{
S~=&~{1\over 8\pi} \int d^2x \big( \del\phi\bar{\del}\phi +
\del X \bar{\del} X \big) -{1\over 4\pi\sqrt{2k'}} \int
d^2x \sqrt g R^{(2)} \phi \cr
& + {\mu\over 8\pi} \int d^2x \bigl( \sqrt{k'\over 2}\del\phi
+ i{\sqrt{k\over 2}}\del X \bigr)\bigl( \sqrt{k'\over 2}\bar{\del}\phi
+ i{\sqrt{k\over 2}}\bar{\del} X \bigr) \exp \big( -\sqrt{2\over k'}\phi
\big)~.\cr}
 \eqno{(43)}
$$
%%%%%%%%%%%%%%%%%%%%%%%%%%%%%%%%%%%%%%%%%%%%%%%%%%%%%%%%%%%%%%%%%%%
(We drop the prime on $\phi,X$ hereafter.)
(43) gives a non-linear $\sigma$-model with the target-space metric
%%%%%%%%%%%%%%%%%%%%%%%%%%%%%%%%%%%%%%%%%%%%%%%%%%%%%%%%%%%
$$
\eqalign{
G_{\phi\phi}~=&~ 1 + {k'\over2}\mu \exp \big(
-\sqrt{2\over k'}\phi \big)~, \quad
G_{XX}~=~ 1 -  {k \over 2}\mu \exp \big(
-\sqrt{2\over k'}\phi \big)~, \cr
G_{X \phi}~=&~i\sqrt{k'\over 2} \sqrt{k\over2}\mu \exp \big(
-\sqrt{2\over k'}\phi \big)~, \cr}
\eqno{(44)}
$$
%%%%%%%%%%%%%%%%%%%%%%%%%%%%%%%%%%%%%%%%%%%%%%%%%%%%%%%%%%%%%
together with the linear dilaton field
%%%%%%%%%%%%%%%%%%%%%%%%%%%%%%%%%%%%%%%%%%%%%%%%%%%%%%%%%%%%%
$$
\Phi = \sqrt{2\over k'} \phi~. \eqno{(45)}
$$
%%%%%%%%%%%%%%%%%%%%%%%%%%%%%%%%%%%%%%%%%%%%%%%%%%%%%%%%%%%%%5
It is easy to check that (44),(45) satisfy the equation of
the vanishing one-loop $\beta$-function,
%%%%%%%%%%%%%%%%%%%%%%%%%%%%%%%%%%%%%%%%%%%%%%%%%%%%%%%%%%%%
$$
R_{ab}= D_a D_b \Phi~, \quad a,b = \phi, X
\eqno{(46)}
$$
%%%%%%%%%%%%%%%%%%%%%%%%%%%%%%%%%%%%%%%%%%%%%%%%%%%%%%%%%%%%
in the leading order in $1/k$. Thus the mass perturbation (42)
is in fact a marginal perturbation of the black hole theory.
\par
In the leading order in $1/k$ the metric (44) is given by
%%%%%%%%%%%%%%%%%%%%%%%%%%%%%%%%%%%%%%%%%%%%%%%%%%%%%%%%%%%%
$$
\eqalign{
G_{\phi\phi}~=&~ 1 + {k \over 2}\mu \exp \big(
-\sqrt{2\over k}\phi \big)~, \quad
G_{XX}~=~ 1 -  {k \over 2}\mu \exp \big(
-\sqrt{2\over k}\phi \big)~, \cr
G_{X \phi}~=&~i{k\over 2} \mu \exp \big(
-\sqrt{2\over k}\phi \big)~. \cr}
\eqno{(47)}
$$
%%%%%%%%%%%%%%%%%%%%%%%%%%%%%%%%%%%%%%%%%%%%%%%%%%%%%%%%%%%%%%%%
After diagonalization $\bigl( d\psi = dX + i k\mu/2 \exp\big(
-\sqrt{2/k}\phi \big)\bigl( 1 - k\mu/2 \exp\big(
-\sqrt{2/k}\phi \big)\bigr)^{-1} d\phi \bigr)$ the metric (47) becomes the
``two-dimensional Schwarzschild metric'' discovered in
\REF\TATA{G.Mandal, A.M.Sengupta and S. Wadia, \mpl{6} (1991)
1685.}
[\TATA],
%%%%%%%%%%%%%%%%%%%%%%%%%%%%%%%%%%%%%%%%%%%%%%%%%%%%%%%%%%%
$$
ds^2~=~\bigl( 1 - {k\mu\over2} \exp\big(-\sqrt{2\over k}\phi
\big) \bigr)^{-1} d\phi^2 + \bigl( 1 - {k\mu\over2}\exp\big(
-\sqrt{2\over k}\phi \big) \bigr) d\psi^2~.
\eqno{(48)}
$$
%%%%%%%%%%%%%%%%%%%%%%%%%%%%%%%%%%%%%%%%%%%%%%%%%%%%%%%%%%%
Its scalar curvature is given by
%%%%%%%%%%%%%%%%%%%%%%%%%%%%%%%%%%%%%%%%%%%%%%%%%%%%%%%%%%%%%%
$$
R = \mu \exp\big( -\sqrt{2\over k}\phi \big)~,
\eqno{(49)}
$$
%%%%%%%%%%%%%%%%%%%%%%%%%%%%%%%%%%%%%%%%%%%%%%%%%%%%%%%%%%
and thus the curvature singularity exists at $\phi = -\infty$.
We also note that
the event horizon occurs at $\phi = \phi^* = \sqrt{k/2}
\ln {\mu k/2}$. Thus the region $\phi^* < \phi < \infty$
describes the asymptotically flat region of the
black hole, while $-\infty < \phi < \phi^*$ describes
the region between the event horizon and singularity.
We note that if one replaces $k$ by $k'=k-2$ in (44) (as in ref. [\WITA])
, it becomes equivalent to (47) and becomes a black hole
in all orders in $1/k$.
The coordinate transformation which brings
the action (43) into that of the gauged WZW model is also easy
to work out. By a simple change of variables [\BEKU]
%%%%%%%%%%%%%%%%%%%%%%%%%%%%%%%%%%%%%%%%%%%%%%%%%%%%%%%%%%%%%%
$$
\eqalign{
\phi &={1\over\sqrt2} log\cosh r + \phi^*~, \cr
\theta &= X + i log\tanh r~. \cr}
\eqno{(50)}
$$
%%%%%%%%%%%%%%%%%%%%%%%%%%%%%%%%%%%%%%%%%%%%%%%
(43) is transformed into
%%%%%%%%%%%%%%%%%%%%%%%%%%%%%%%%%%%%%%%%%%%%%%%
$$
S_{WZW}~=~{k\over 4\pi} \int d^2x \bigl( \del r \bar{\del} r
+ \tanh^2 r \del\theta\bar{\del}\theta \bigr)
-{1\over 4\pi} \int d^2x \sqrt g R^{(2)} log\cosh r~,
\eqno{(51)}
$$
%%%%%%%%%%%%%%%%%%%%%%%%%%%%%%%%%%%%%%%%%%%%%%%%%%%%%%%%%%%%%
in the leading order in $1/k$.
(51) is the effective action of the gauged WZW model obtained
after eliminating the gauge field [\WITA].
The presence of $i$ (imaginary unit)
in formulas (44) (50) is somewhat disturbing. However, it disappears
when we go to the Minkowski convention $(X \to iX)$ and there the
change of variables seems well-defined.
\par
We should remark that our results on the physical spectrum of the coset model
$SL(2;R)/U(1)$ do not agree that those obtained by Distler and Nelson
\REF\DINE{J. Distler and P. Nelson, \np{374} (1992) 123.}
[\DINE] based on a different method for quotienting the $U(1)$ symmetry.
Extra physical states found in [\DINE] in the coset model become
BRST trivial states in our formulation.
\par
In this paper we have considered the cohomology $H_{Q}$
of the total BRST operator $Q = Q_{diff}+Q_{U(1)}$. We believe
that the non-zero cohomology of $Q_{U(1)}$ is concentrated
at the ghost number zero sector and hence $H_{Q}$ agrees with
that of $H_{Q_{diff}}(H_{Q_{U(1)}})$ where $Q_{diff}$ cohomology is
taken after computing the $Q_{U(1)}$ cohomology.
\par
T.E. would like to thank Profs. M.Atiyah and P.Goddard for their hospitality
at Newton Institute for Mathematical Sciences, University of Cambridge.
\par
The research of T.E. and S.K.Yang is partly supported by Grant-in-Aid for
Scientific Research on Priority Area ``Infinite Alalysis''.
\refout

\end